\newcommand{\simle}{\mbox{$\stackrel{<}{_{\sim}}$}}
\newcommand{\simge}{\mbox{$\stackrel{>}{_{\sim}}$}}

\newcommand{\micronn}{\,\hbox{$\mu$m}}
\def\etal{ {\em et~al.\/}\thinspace}
\def\arcsec{\hbox{$^{\prime\prime}$} }

\def\kms{ km~s$^{-1}$}
\def\rstar{\,R$_\star$}
\def\wn{\,cm$^{-1}$}

\documentstyle[11pt,aaspp4,epsfig,flushrt]{article}
\lefthead{Monnier et al.}
\righthead{Mid-IR interferometry on Spectral Lines}
\begin{document}

\title{Mid-infrared interferometry on spectral lines: 
\\ 
III. Ammonia and Silane around IRC~+10216 and VY~CMa}

\author{J. D. Monnier\altaffilmark{1}, W. C. Danchi\altaffilmark{2},
D. S. Hale, P. G. Tuthill\altaffilmark{3}, C. H. Townes}

\altaffiltext{1}{Current Address: Smithsonian Astrophysical Observatory MS\#42,
60 Garden Street, Cambridge, MA, 02138}
\altaffiltext{2}{Current Address: NASA Goddard Space Flight Center,
Infrared Astrophysics, Code 685, Greenbelt, MD 20771}
\altaffiltext{3}{Current Address: Chatterton Astronomy Dept, 
School of Physics, University of Sydney, NSW 2006, Australia}
\affil{Space Sciences Laboratory, University of California, Berkeley,
Berkeley,  CA  94720-7450}

%
%
%
\begin{abstract}
Using the U.C. Berkeley Infrared Spatial
Interferometer with an RF filterbank, the first
interferometric observations of mid-infrared 
molecular absorption features of ammonia (NH$_3$)
and silane (SiH$_4$) with very high spectral resolution
($\frac{\lambda}{\Delta\lambda}\sim10^5$) were made.  
Under the assumptions of spherical symmetry and uniform outflow, 
these new data permitted
the molecular stratification around carbon star IRC~+10216 and red
supergiant VY~CMa to be investigated.  For IRC~+10216, 
both ammonia and silane were found to form in the dusty outflow
significantly beyond both the dust formation and gas acceleration
zones.  Specifically, ammonia was found to form
before silane in a region of decaying gas turbulence
($\simge$20\rstar), while the
silane is produced in a region of relatively smooth gas flow much further
from the star ($\simge$80\rstar).  The depletion of gas-phase
SiS onto grains soon after dust formation may fuel silane-producing
reactions on the grain surfaces.  For VY~CMa, a
combination of interferometric and spectral observations suggest
that NH$_3$ is forming near the termination of the gas acceleration
phase in a region of high gas turbulence ($\sim$40\rstar).
\end{abstract}


\pagebreak
\section{Introduction}

Both high and low-mass stars near the end of their lives are known to
emit copious amounts of material with the high mass-loss rates
critically dependent on the formation of silicate or carbonaceous dust
grains within a few stellar radii of the red (super-) giant
photosphere.  Stellar photons both impart momentum to the dust
particles through absorption and scattering (which drive a wind) and
heat dust close to the star to temperatures up to 1200-1500\,K
(e.g., \cite{lafon91}).  
Thermal radiation from these hot dust grains peaks in the near-infrared
(1-5\micronn), while more distant and 
cooler (500\,K) grains emit mostly in the
mid-infrared (5-20\micronn).

Although the physical sizes of these evolved stars are impressive
(R$_\star \simge 1$ AU), their galactic paucity means that even the
closest examples lie at distances greater than 50~pc. Thus the
characteristic size scale for dusty circumstellar emission at a few stellar
radii is
generally a small fraction of an arcsecond,
too small to be resolved using standard observing techniques
from ground-based telescopes limited by telescope diffraction (in
the mid-infrared) and/or atmospheric turbulence (for shorter wavelengths).
However, long-baseline interferometry in the infrared can
directly detect and measure the morphologies of these envelopes.

The high densities of heavy elements and the mild temperatures around
evolved stars encourage 
the formation of myriad diatomic and polyatomic
molecules in addition to dust grains.  
For instance, over 50 molecular species have been found
around prototypical carbon star IRC~+10216
(see review by Glassgold [1998]).  Attempts to
understand the density, temperature, and velocity distributions as well
as the formation mechanisms of
these molecules have contributed to the development of the field of
astrochemistry.

Using estimates of the temperature, atomic abundances, and gas density
in and above the photosphere, predictions of molecular abundances can
be made through detailed calculations of a network of chemical
reactions.  By assuming local thermodynamic equilibrium (LTE), various
molecules are said to ``freeze-out'' in the stellar wind.  The
decreasing temperatures and densities cause reaction rates to fall
quickly, locking the atoms in certain energetically favorable
molecules.  Frozen equilibria models have proven useful at explaining
the abundances of many molecules observed around AGB stars (e.g.,
\cite{lafont82}), however additional chemical processes are needed to
explain some detected species.  For instance, penetration of
interstellar UV-radiation catalyzes photochemical reactions in the
outer envelopes of AGB stars (e.g., \cite{cg93}),  while other
molecules may be produced under non-equilibrium conditions associated
with shocks (\cite{wc98}).  Particularly relevant here is the fact
that negligible amounts of silane and ammonia are predicted by
most freeze-out models, yet they have been detected with some abundance
in circumstellar envelopes (e.g., \cite{keady93}); their formation is
hypothesized to be catalyzed on the surfaces of dust grains in the flow.
In addition, 
recent discoveries of brown dwarfs and extrasolar planets have catalyzed
interest in the chemistry of cool stellar atmospheres (e.g., 
Burrows \& Sharp 1999), leading to new insights into molecular formation
mechanisms in these environments.  Since it is often
difficult or impossible to experimentally reproduce the physical
conditions of circumstellar and interstellar space, molecular
observations are critical to test and guide relevant theories.

This paper is the third in a series 
on the spatial distributions of
dust and molecules in the inner envelopes of nearby red giants and
supergiants.  Paper I (\cite{monnier2000a}) discussed the hardware 
implementation of this experiment, while
Paper II (\cite{monnier2000b}) discussed recent
visibility data at 11.15\,$\mu$m  
and presented the appropriate
dust shell models for IRC~+10216 and VY~CMa.
This paper (Paper III) makes use of these results as well as the first
mid-infrared
visibility data ever reported on spectral lines, providing critical new 
information on the molecular stratification around these stars.

The methods used in this work
are technologically challenging and represent
significant advances in infrared interferometry.  By combining the
high spatial resolution of long-baseline interferometry with the high
spectral resolution of heterodyne spectroscopy (Paper I),
this work has, for the first time, probed the brightness distribution
on and off of spectral absorption features as narrow as $\sim$1\kms
($\frac{\lambda}{\Delta\lambda}\sim10^5$).  This allows the absorbing
regions of polyatomic molecules to be directly measured,
setting strong limits on the formation radii.  Such measurements are
important for determining the formation mechanism since there
is presently no good theory to explain the observed
high abundance of certain
molecular species (e.g., NH$_3$ and SiH$_4$).

\section{Modeling}
The observing methodology for extracting visibility data on and
off of spectral lines with the Infrared Spatial Interferometer
has been discussed in detail in Paper~I and Monnier (1999).
New dust shell models
of IRC\,+10216 and VY CMa have been developed in Paper II.
However, interpreting spectral line observations
requires modeling of the distribution of circumstellar
molecules as well as dust.

As for observations of
continuum radiation with the ISI, 
it is rarely possible to directly reconstruct
an image of the astrophysical source due to the sparse sampling of the Fourier
plane afforded by a two-element interferometer.  
This limitation
necessitates the use of radiative transfer models for
interpreting the line data.  This section discusses the 
assumptions used in creating models of the molecular
envelopes around AGB stars, the numerical code, and the scientific goals
of subsequent analyses. 

\subsection{Simple models}
In Paper~II, the assumptions of a spherical symmetric and uniform
outflow of dust embedded in the stellar winds were adopted,
approximations which have been extended to include the distribution of
molecular material.  As with the models of the dust shells, the
molecular envelope was characterized by the inner formation radius and
overall abundance factor, free parameters in fitting models to the
molecular line data.

Observations have shown that the line shapes of SiH$_4$ and NH$_3$
around AGB stars typically show emission in the red wing and
absorption in the blue (e.g., \cite{goldhaber88}).  This kind of
profile is expected for molecular emission and absorption in an
expanding envelope (e.g., \cite{shu91}).
Figure\,\ref{fig:models_line} shows a theoretical line profile similar
to those actually observed.  The bulk of the absorption is
blue-shifted with respect to the star at the speed of the outflow.
This is because the entire column of molecular gas along the
line-of-sight connecting the observer and the star is at the same
relative velocity (-v$_{\rm outflow}$), assuming uniform outflow;
hence, there is a large optical depth, $\tau_{\rm line}$, at this
frequency and self-absorption is observed.  
On the other hand,
lines-of-sight which penetrate the molecular gas envelope at impact
parameters off the observer-star axis encounter parcels of gas at a
variety of relative doppler shifts.  The optical depth at any
particular frequency is hence generally low, allowing emission from
deeper layers of hot dust to escape and to be observed without
significant molecular absorption.  The low-level emission observed at
both blue- and red-shifted velocities arises from relaxation of
IR-pumped and collisionally-excited transitions into higher energy
vibrational states.  Note that absorption in the line core 
occurs primarily because of the large optical depth at that frequency 
combined with the fact that the temperature 
decreases with increasing distance from the star.
Hence, absorption under these conditions would occur
even when the molecules are in local thermodynamic equilibrium (LTE)
with the dust and radiation.  

For the AGB stars under study here, 
the source of most of the mid-infrared flux is thermal emission from
the dust envelope (\cite{danchi94}), not from the star
itself.  
Since the blue-shifted absorption features indicate that most
of the absorption is occurring along the direction towards the center
of dust envelope, an interesting measurement that can be made with an
interferometer is the size of the absorption region.  This measurement
yields direct information on the approximate location of the
$\tau_{\rm line}\sim 1$ surface, which can be used to determine the
abundance and inner radius of molecular formation.
However this method is insensitive to molecules which may exist
right at the inner radius of the dust, if they are in LTE.  This is
because there is no significant background source of radiation to absorb.
These complications require the use of a sophisticated spectral line code
to quantify the sensitivity of this method to molecules forming 
near the inner radius of the dust shell.  Such a code has been used, 
and is described in the next section.

Previous spectral line work (\cite{betz79}; \cite{mclaren80}; 
\cite{goldhaber84};
\cite{goldhaber88}; \cite{keady93}) has shown
that ammonia and silane exist around AGB stars at up to 10$^4$ times
the abundances predicted by equilibrium calculations of 
expanding outflows (for more recent work under somewhat different
physical conditions, see Burrows \& Sharp 1999).
It has been suggested
that dust catalyzes the formation of these
molecules through chemical reactions on grain surfaces.  
If this is the case, 
one
might expect the inner radius of molecular formation
to coincide with the dust formation radius.  However, atoms and
molecules generally have low sticking efficiencies at high
temperatures, elastically scattering off the grain surfaces (e.g.,
\cite{sticky85}). This fact and the lack of 
known chemical pathways for the formation of molecules such as ammonia and
silane make direct
measurements of molecular formation radii important, and this
is one of the prime goals of the filterbank
project of the Infrared Spatial Interferometer.

\subsection{Radiative transfer calculations} 
Radiative transfer in spectral lines often is accomplished by applying the
Sobolev approximation (e.g., \cite{goldhaber88}; \cite{shu91}), 
which assumes that
large-scale velocity gradients essentially radiatively
decouple parcels of molecular
gas.  To accurately calculate the radiative transfer in a spectral line, the
Sobolev approximation requires the natural linewidth, in this case determined
by microturbulence (for cold gas) or thermal motions (for hot gas), to be
much smaller than the expansion velocity.  While this condition is reasonably
satisfied for the narrowest lines ($v_{\rm outflow}\sim15\times \Delta v_{\rm
linewidth}$), it is only weakly so for the broadest ones ($v_{\rm
outflow}\sim4\times \Delta v_{\rm linewidth}$).  

In order to avoid this
uncertainty, all radiative transfer calculations in spectral lines were
performed using a code developed by Dr. J. J. Keady (\cite{keady82}), based on
the method of Mihalas\etal (1975).  It accurately calculates line
profiles as well as frequency-dependent emission profiles, from which
visibility curves can be computed on and off spectral features.  This
code was developed to treat the case of molecules embedded in an expanding
flow, even when the absorbing molecules are comingled with 
continuum-emitting sources. This is likely the case for AGB stars where
thermal emission by dust is believed to be ``filling-in'' the absorption
features.  The program, written in Fortran, formally solves the observer's
frame transport equation, which is necessary to correctly account for the
propagation of line radiation through an expanding flow, thus avoiding
the questionable Sobolev approximation.  Dr. Keady has
allowed his code to be used for the analyses which follow, and has
assisted in porting the Fortran code to run under Solaris.  This code has
been extensively tested,
having been used for over
15\,years (\cite{keady82}; \cite{keady88}; \cite{keady93}; \cite{winters98}).
While the code was already able to be used with
spherical top molecules (e.g., SiH$_4$), the appropriate partition function
and statistical weights had to be programmed for NH$_3$ (a symmetric rotor).
A nice summary of the mid-infrared molecular properties of both SiH$_4$ and
NH$_3$ can be found in chapter 5 of Goldhaber's thesis (1988).  Readers
interested in the numerical details of the radiative transfer calculation
should consult Keady (1982)  and
Mihalas\etal (1975) for further details.

\subsection{Assumptions}
\subsubsection{Dust}
The source function used for radiative transfer in Keady's code assumes a
single dust grain size and temperature at a given radius from the star.  The
code (\cite{wolfire86}) 
which was used for fitting the ISI continuum visibility data in Paper II
utilized the full
MRN (\cite{mrn}) 
distribution of grain sizes and hence some modification must be made to adopt
our previous models for
use by Keady's algorithm.  The average cross-section of the MRN
distribution was used as input to the Keady code, along with the 
size-averaged dust
temperature. 
This necessary simplification resulted in a slight misfit to the ISI
continuum visibility data, which was compensated for by empirically adjusting
the overall value of the single grain-size dust opacity ($\simle$30\%
change).

\subsubsection{Temperature(s) of the gas}
\label{section:gastemps}
In addition to spherical symmetry and uniform outflow, other
assumptions are also generally made for line calculations.  Most
important is the assumption that the occupation of various
ro-vibrational states of the polyatomic molecules can be approximated
by a Boltzmann distribution using separate vibrational and rotational
temperatures, T$_{\rm vib}$ and T$_{\rm rot}$.  Ideally, 
the populations of the various states should be calculated 
using a multi-level
model molecule.  However, the collisional excitation constants are
poorly known for SiH$_4$ and NH$_3$ and the large number of (far-IR)
rotational transitions make for a difficult and uncertain result.

Goldhaber (pg.103, Goldhaber [1988]) considered this question in some
detail for IRC\,+10216 and his conclusions will now be summarized.
For sufficiently high molecular densities, both mid-IR and far-IR
(i.e., ro-vibrational and pure rotational) transitions will be
optically thick, trapping line radiation. In this case, collisional
excitation and de-excitation will have a strong influence on the level
populations, equilibrating the vibrational and rotational temperatures
with the gas (kinetic) temperature.  However, the molecules under
consideration here are not found at sufficiently high densities for
this to apply.

At the somewhat lower densities encountered for ammonia and silane, the
mid-IR (ro-vibrational) transitions become optically thin.  In terms of
excitation mechanism, this allows IR pumping of the vibration-rotation
transitions to dominate (over collisions) within about 70\,R$_\star$, while
rotational relaxation in the pure rotational transitions (far-IR) dominate in
the outer envelope (where these transitions become optically thin). 
Collisional excitation and de-excitation never play a dominant role.  Hence,
T$_{\rm vib}$ and T$_{\rm rot}$ should be close to the radiation/dust
temperature, T$_{\rm dust}$, suggesting that 
T$_{\rm vib}$=T$_{\rm rot}$=T$_{\rm dust}$
is a good starting point. The radial dependence of the rotational
and vibrational temperatures usually need slight adjustment during the
modeling process to reproduce the relative depths of high and low excitation
lines.

One interesting exception applies to the J=K states of NH$_3$.  The
rotational dipole moment along the symmetry axis (z-axis) of NH$_3$ is equal
to zero; hence the $\Delta K = 0$ selection rule must apply for all radiative
(dipole) transitions (see Townes \& Schawlow [1975]).  Therefore, a
molecule rotating in the J=K state can not radiatively de-excite to a lower
energy (lower J) state.  Hence, the equilibrium population of J=K states is
{\em not} determined by the radiation temperature but by collisions with the
ambient gas (mostly H$_2$); the rotational temperature of these states should
be in equilibrium with the gas (kinetic) temperature.  However,
a strong (parallel-type) vibrational band at 6.1\,$\mu$m does
allow $\Delta K=\pm 1$ transitions, thereby allowing K-ladders to come into
equilibrium if sufficiently excited by the radiation field.
Sufficient IR pumping in this band is likely to dominate over
collisional processes out to a few hundred
\rstar, although a detailed calculation is lacking
(see Goldhaber [1988]).

\subsubsection{Molecular constants}
The polyatomic molecules of SiH$_4$ and NH$_3$ have been
modelled using physical constants, including the moments of inertia and
vibrational energy levels, adopted from Keady \& Ridgway (1993) and
Goldhaber (1988).  Readers interested in learning more
about molecular spectroscopy should consult these sources; 
a rigorous and complete treatment of the subject can be found in {\em Microwave
Spectroscopy} by Townes \& Schawlow (1975).

\begin{table}[htb]
\caption[Molecular constants]{\scriptsize This table
lists the molecular constants assumed in the model
calculations.  Most of these parameters were originally
collected in Goldhaber (table V.1 in Goldhaber [1988]).
The rotational constants (B \& C) are given for the
vibrational ground state along with the appropriate
vibrational dipole moment ($\mu_n$) and
origin frequency ($\nu_n$).
\label{table:mol_constants}}
\begin{center}
\begin{tabular}{ll}
\hline\hline
{\bf SiH$_4$} & \\
& \\
B = 2.8591\wn & Gray, Robiette, \& Johns (1977) \\
$\nu_4$ = 913.473\wn & Gray, Robiette, \& Johns (1977) \\
$\mu_4$ = 0.247 debye & Keady \& Ridgway (1993) \\
\hline
{\bf NH$_3$} & \\
& \\
B = 9.9466\wn & Urban\etal (1983) \\
C-B = -3.7199\wn & Urban\etal (1983) \\
$\nu_2$ = 932.434\wn & Urban\etal (1983) \\
$\mu_2$ = 0.24 debye & Nakanaga, Kondo, \& Saeki (1985) \\
\hline
\end{tabular}
\end{center}
\end{table}

\subsection{Analysis method}

\label{subsection:goals}
For each star and target molecule, 
the analysis strategy for ISI spectral line data was straightforward and
had three major steps.

\begin{enumerate}
\item{The dust shell was modelled in order to have a 
reasonable approximation of the continuum source.  This has been done in
Paper~II, where coeval continuum visibility
data were used to create appropriate
dust shell models of the astrophysical sources
under study.}
\item{Molecules were assumed to form at a given radius from the star
and flow out uniformly.
Theoretical
spectral line profiles were then calculated to fit both ISI measurements
and also previous observations of spectral line depths and ratios.  
Lines of various
excitation energies were used (when available)
in order to probe the abundance, temperature,
and turbulent
structure of the gas.  This level of analysis relied heavily on the careful
modeling pre-existing in the astrophysics literature.}
\item{Theoretical visibilities on and off the spectral line were calculated 
and
compared to the observed visibility ratios.  
The radius of molecular formation set in step~2 
was varied until all available data were satisfactorily
fit.}
\end{enumerate}

This method allows one to distinguish between gas models with high
molecular abundance far from the star and those with lower molecular abundance
close to the star in two different ways.
While the line depth of a given transition can always be fit by 
varying the abundance and mean gas temperature, the {\rm relative} 
line depths of transitions with different excitation energies
can not.  This probe of the average excitation temperature, when coupled
with estimates of the radial temperature profile and assumption of
uniform outflow, can be used to infer
the location of molecular formation (e.g., \cite{keady93}).
Here in this work, we also apply a second, more direct, approach by using
an interferometer.  With suitable spectral and spatial resolution, one
can directly detect the 
absorption pattern when the molecules are sufficient close to the star. 
This is the main thrust of this paper, where we show that molecules
within 40\rstar would significantly modify the emission pattern in the line 
core, providing a method independent of the line ratio argument for
detecting the presence of molecules close to the star.

\subsection{Comparing model quantities to observations}
After fully specifying the dust and gas characteristics, the Keady
code returns a spectral line profile and monochromatic radial emission
profiles.  This section discusses some of the details regarding
comparison of the modelling results with ISI filterbank data.

\subsubsection{Spectral line profiles}
Because of the heterodyne detection scheme,
mid-infrared radiation arriving outside of the primary beam of the 
interferometer was not seen.  This had a small effect on the depth
of the spectral line, since only a fraction of the 
molecular absorption occurs
in the outer envelope.  The effect was compensated for by multiplying the
angular distribution resulting from 
monochromatic radial emission calculations by the effective
primary beam of FWHM~3\,\arcsec (see Paper II for more discussion)
and then adjusting the flux level accordingly.  All spectral profiles 
shown in the next sections have had this correction applied.  

Next, the finite bandwidth of the filterbank observations and the
double-sideband (DSB) nature of the detection were accounted for.  The
bandpass center frequency and width were used to determine the average
depth of the calculated spectral line for a corresponding
single-sideband (SSB) observation.  The dilution of the line depth due
to the combination of the ``uninteresting'' additional continuum
sideband encountered in practical observation was accounted for by
dividing the SSB line depth by 2.  This resulting DSB (diluted) line
depth was then compared to observations with the filterbank.

\begin{figure}[hbt]
\begin{center}
\epsfig{file=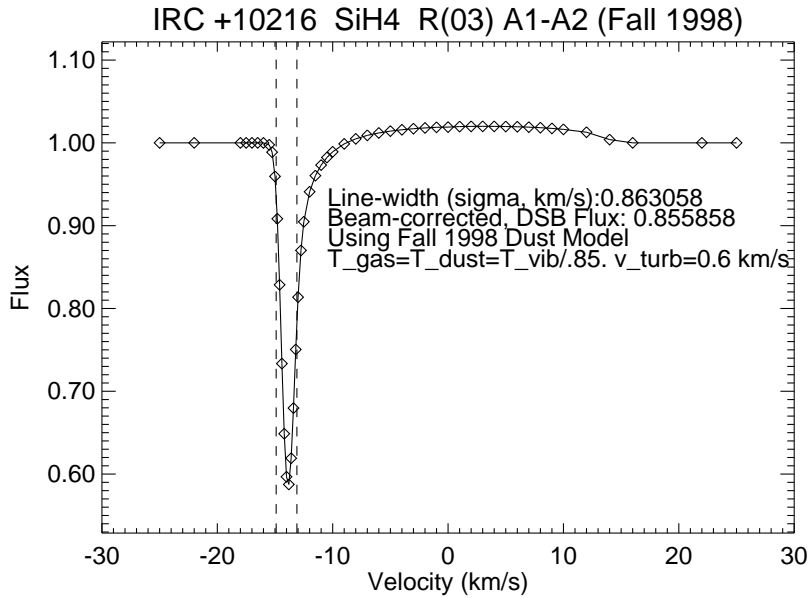,width=4.5in}
\caption{Example of
a theoretical spectral line shape using radiative
transfer calculations
from the Keady code (\cite{keady82}).
\label{fig:models_line}}
\end{center}
\end{figure}

Figure\,\ref{fig:models_line} shows one of the output figures from the
modeling analysis suite.  The bandwidth of the observation is marked off by
vertical dashed lines, and line diagnostics have been calculated.
The line width and average DSB line depth were also calculated for
comparison with spectral line observations.  Other quantities
were also determined, such as the equivalent width, which were
useful when comparing with lower spectral resolution data.

\subsubsection{Visibility ratios}
The basic observable of the filterbank
experiment is the visibility ratio on the spectral line compared to
the dust continuum.  Monochromatic radial emission profiles were 
calculated at 10 specific frequencies across the spectral line, sampling
the stellar continuum as well as the entire absorption region.
The double-sideband, radial emission profile for the finite bandwidth selected
by the filterbank was determined
by a flux-weighted average of the emission profiles inside the
observation bandwidth and an equal bandwidth of continuum.  This
resultant profile was multiplied by the effective primary beam and
the visibility then calculated.  The visibility curve of the
continuum emission was calculated as well, and
the visibility ratio (on the line compared to the continuum)
as a function of baseline was then determined.

\begin{figure}[htb]
\begin{center}
\epsfig{file=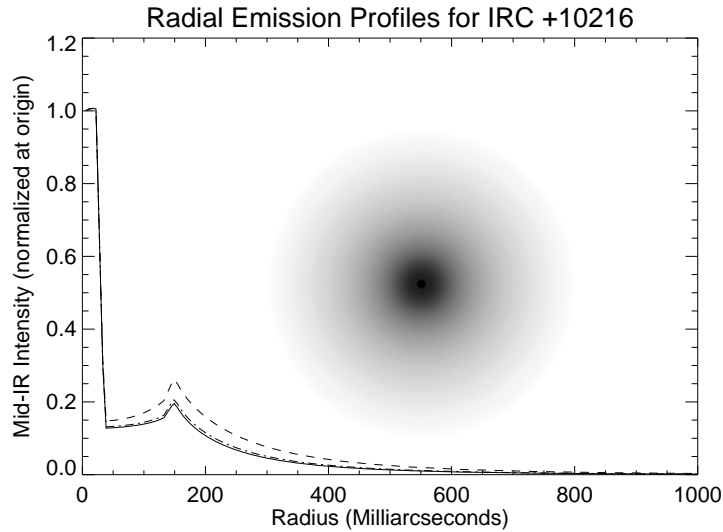,width=4.0in}
\caption{\scriptsize This figure shows an example of the mid-IR emission as a function 
of radius from the star for one particular gas model, normalized
to unity at radius = 0.  The solid line shows
the radial profile for the dust continuum, while the dashed line
represents the emission at the core of the absorption line.
The dot-dashed line represents the emission
after accounting for the finite bandwidth and double-sideband detection.
The inset image
is a two-dimensional representation of the
absorption region defined by the ratio of the absorption core to the continuum
profiles.
\label{fig:models_profiles}}
\end{center}
\end{figure}

\begin{figure}[hbt]
\begin{center}
\epsfig{file=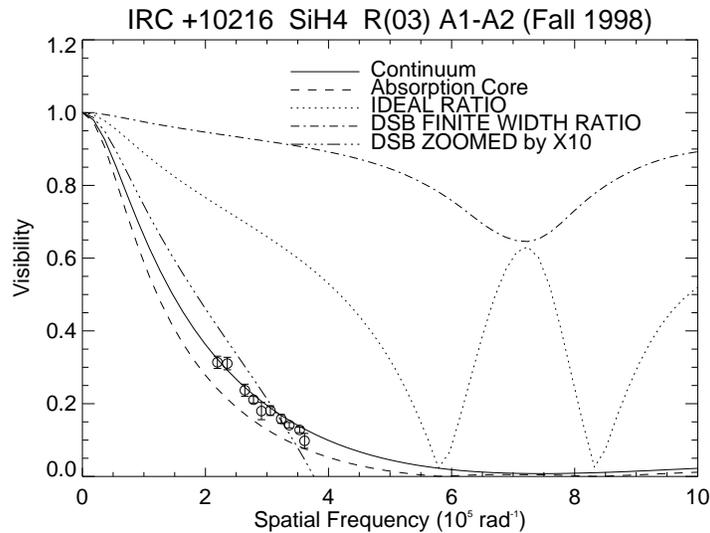,width=4.0in}
\caption{\scriptsize
This figure shows visibility curves on and off a spectral feature.
The dashed and solid lines represent the visibility curves on and off the
absorption core respectively; the dotted line shows the ratio.
The dot-dashed line shows the visibility ratio after accounting for
finite bandwidth and double-sideband (DSB) detection.  Deviations from
unity are multiplied, or zoomed in, by a factor of 10, and shown with a
3-dots/dashed linestyle; the zoom means that, for example,
0.6 now corresponds to 0.96 in
the original units.  The data points on this plot represent coeval
visibility measurements of the dust continuum reported in Paper II.
\label{fig:models_vis}}
\end{center}
\end{figure}

Figures\,\ref{fig:models_line}-\ref{fig:models_vis} illustrate this
entire process, showing example output at every stage of the analysis
process for a silane line around IRC~+10216.  All subsequent model
visibility ratios have been processed in identical fashion, but these
diagnostic plots are not presented for brevity.  The particular gas
model shown in figures\,\ref{fig:models_line}-\ref{fig:models_vis} has
a silane formation radius of only 10\,R$_\star$, small enough to
produce a large change in the visibility on and off the spectral line.
Since the filterbank system measures a visibility ratio, the radial
emission profiles were usually of only secondary interest, and more
emphasis was placed on interpretations of ratios.

To facilitate a better understanding of the visibility data, example
brightness profiles appear in figure\,\ref{fig:models_profiles}.  The
bright spike at the origin is from stellar photospheric emission seen
through the dust and gas, while the secondary peak at 150\,mas reveals
the location of the dust shell inner radius.  It is important to note
that while the star itself clearly shines through the dust envelope
(it is the highest surface brightness feature of the nebula), it
contributes a tiny percentage of the total mid-IR emission due to the
large size of the dust shell.  The inset image is a two-dimensional
representation of the absorption region defined by the ratio of the
absorption core to the continuum profiles.  One can see that
absorption for this model
is indeed highest near the center of the dust shell.  The
basic point of the filterbank experiment was to measure the size of
this absorption region.

Figure\,\ref{fig:models_vis} shows the visibility curves corresponding to
the emission profiles of figure\,\ref{fig:models_profiles}.  
The various curves are described in the figure caption, and 
show the effects of applying the instrumental
corrections discussed in this section.
In this case, a visibility ratio of about 0.92 would
be expected for the range of baselines observed (2-4\,m).  

The analysis process discussed above took into account all the practical
details of the experiment: finite bandwidth, primary beam, double-sideband
detection. 
The next sections present the data from the filterbank experiment and
detail the modeling results.

\begin{figure}[htbp]
\begin{center}
\centerline{\epsfxsize=\columnwidth{\epsfbox{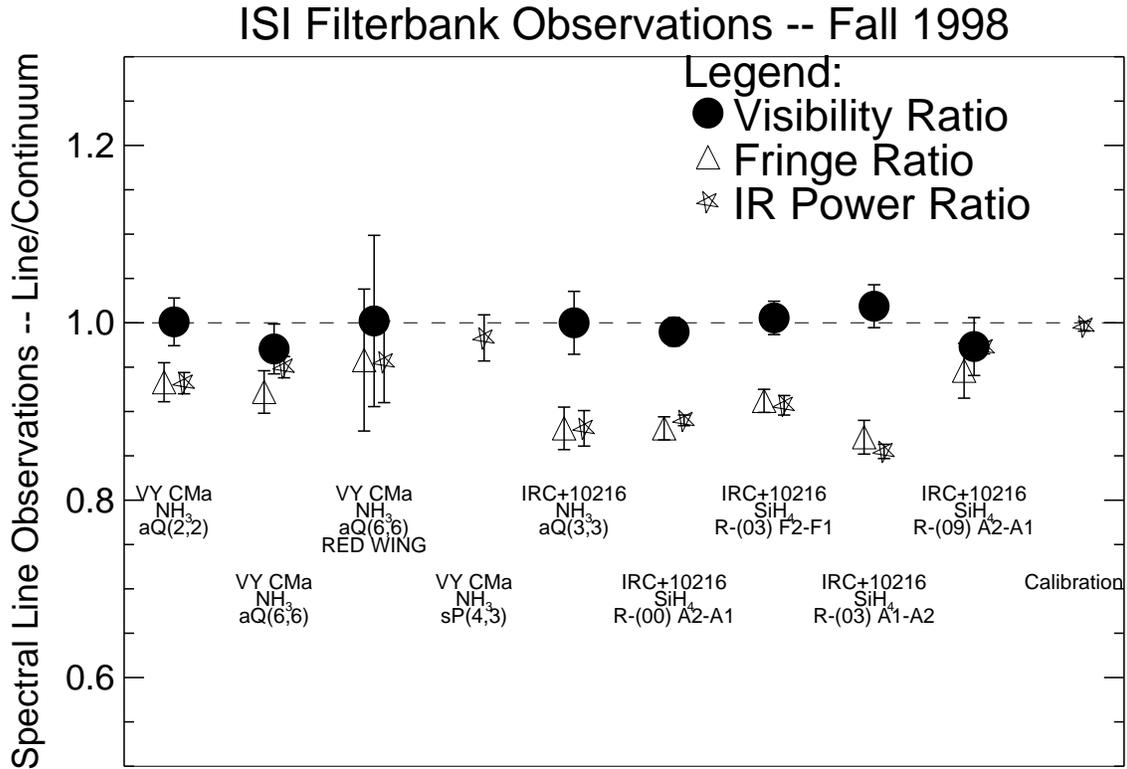}}}
\caption{
This figure is a summary of all the filterbank data on spectral
lines presented in this paper, from 22 full nights of observing in
Fall 1998.  Each data point represents the average of multiple nights
of observing at baselines between 2 and 4 meters, with the
molecular transition and target source indicated below each set of symbols.
Observing details
can be found in tables~\ref{table:10216journal} \&
\ref{table:vycmajournal}.  The triangles show the ratio of the fringe
amplitude on and off the line, while the star symbols show the ratio of
the total infrared (IR) signal on and off the line. The
normalized visibility ratio on and off the spectral line results from
dividing the fringe ratio into the IR power ratio, and appear in the
figure as filled circles.
\label{fig:fb_results}}
\end{center}
\end{figure}

\section{Interferometry on Spectral Lines: Results}
Spectral line observations were carried out for two target sources,
the carbon star IRC~+10216 and the red supergiant VY~CMa.  
General introductions to both of these stars, including
discussions of previous continuum observations, can be found in Paper II.  
A summary of 
all the filterbank data can be found in figure\,\ref{fig:fb_results};
details will be discussed in subsequent sections.  Recall from Paper I that
the ratio of the visibility on the line compared to off the line is made from
dividing the fringe amplitude ratio by the infrared power ratio.  Hence,
in figure\,\ref{fig:fb_results}, the ``Visibility Ratios'' were derived from
the ``Fringe (Amplitude) Ratios'' and the ``IR Power Ratios,'' and do
not represent an independent set of measurements.
See tables of line
frequencies in Monnier (1999, tables C.4 \& C.5) for more information
regarding specific molecular transitions.

\subsection{Comparison with previous spectroscopic results}
\label{section:compare_previous_ir}
\begin{figure}[hbt]
\begin{center}
\centerline{\epsfxsize=5in{\epsfbox{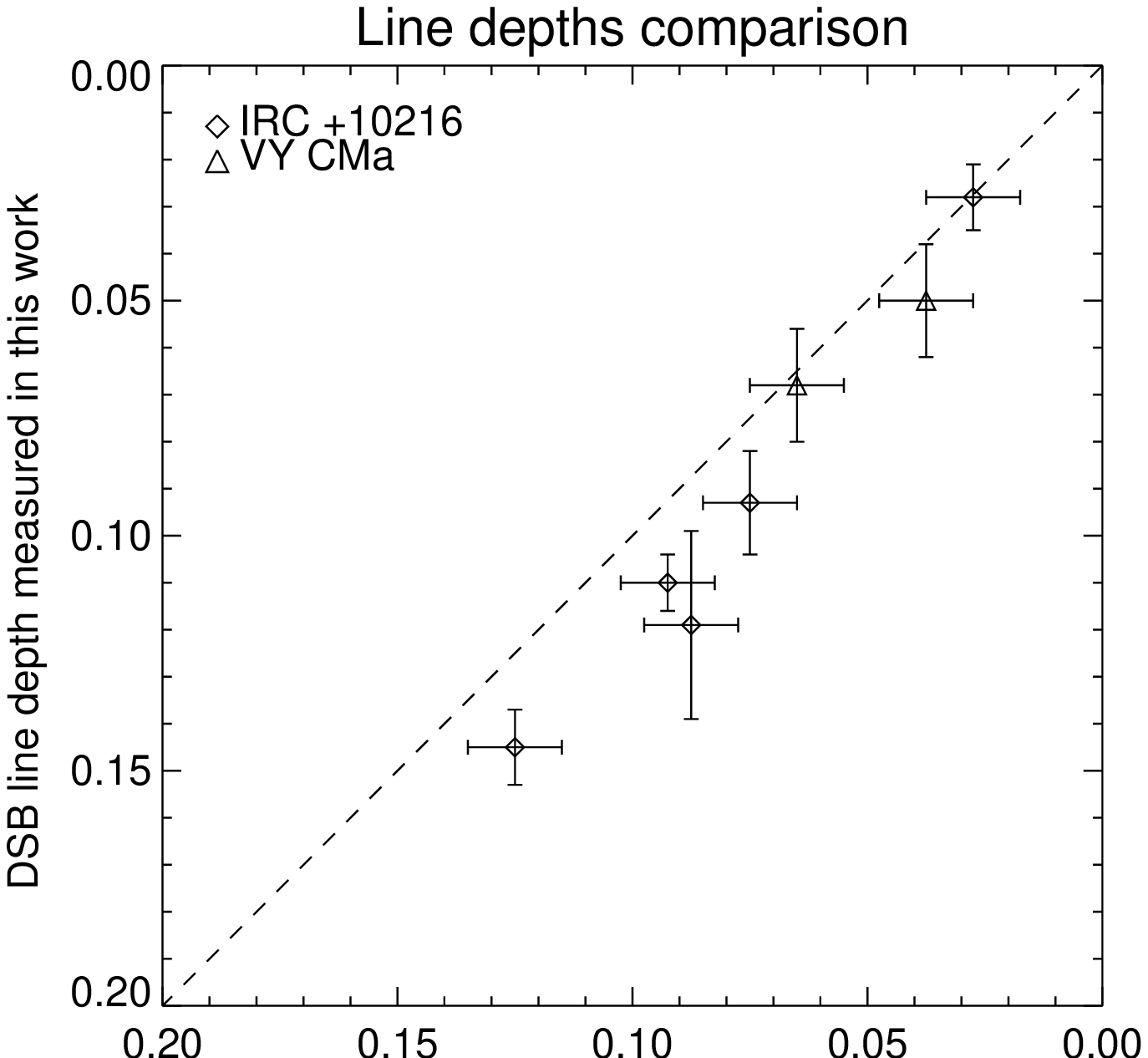}}}
\caption{
This figure compares the
observed line depths of transitions observed by both Goldhaber
(1988)
and this work.  The overall agreement is good, although there is clear
evidence of a systematic difference.  See
\S\ref{section:compare_previous_ir} in
text for discussion.
\label{fig:compare_ir}}
\end{center}
\end{figure}

While the interferometric data presented here are the first of its
kind, the depths of the spectral lines have been observed before.
Figure\,\ref{fig:compare_ir} shows a comparison of all the lines
observed with the double-sideband line depths observed
by Goldhaber (1988).  While the agreement is good in
general, there is a clear signal that the line depths measured in 1998
were slightly deeper around IRC~+10216 than the mid-1980s observations of
Goldhaber.  There are many possible
explanations for this.  Changes in the dust shell geometry,
molecular abundance, as well as different seeing conditions during the
observations themselves can cause small changes in the line depth.
However, the good overall agreement is confirmation
that the circumstellar
environments important for forming these spectral lines
around both VY CMa and IRC~+10216 have not changed radically
in the last decade, justifying the use of previous modeling efforts
(\cite{goldhaber88,keady93})
in developing molecular gas models (see modeling step 2 in
\S\ref{subsection:goals}).

\section{IRC +10216}
\label{section:10216_line_results}
A journal of spectral line observations for IRC~+10216 can be found in
table\,\ref{table:10216journal}, while a full and concise summary of the
data appears in table\,\ref{table:10216data}.  Separate modeling for
both molecules, SiH$_4$ and NH$_3$,  is presented below.

\begin{deluxetable}{cccccl}
\scriptsize
\tablecaption{
Journal of IRC+10216 spectral line observations
\label{table:10216journal}}
\tablehead{
\colhead{Target}        &\colhead{Target}       &
\colhead{LO}    &\colhead{LO}   &\colhead{LO} & \colhead{Dates} \\
\colhead{Molecule}&\colhead{Transition}&\colhead{Molecule}&
\colhead{Transition}&\colhead{Wavelength} & \colhead{(U.T.)}\\
         &          &        &          &\colhead{($\mu$m)} &  }
\startdata
SiH$_4$& R-(00) A2-A1 & $^{13}$C$^{18}$O$_2$ & P(20) & 10.945 &
1998 Nov 05, Nov 06, Nov 13 \\
SiH$_4$& R-(03) F2-F1 & $^{13}$C$^{16}$O$_2$ & R(10) & 10.318 &
1998 Oct 18, Oct 19, Nov 21\\
SiH$_4$& R-(03) A1-A2 & $^{13}$C$^{18}$O$_2$ & P(10) & 10.853 &
1998 Nov 04, Nov 10, Nov 14 \\
SiH$_4$& R-(09) A2-A1 & $^{13}$C$^{16}$O$_2$ & R(30) & 10.694 &
1998 Oct 27, Oct 28, Nov 27, Dec 03, Dec 11 \\
\hline
NH$_3$ & aQ(3,3) & $^{12}$C$^{16}$O$_2$ & P(34) & 10.741 &
1998 Oct 20, Oct 21, Oct 22, Nov 20\\
\enddata
\end{deluxetable}

\begin{deluxetable}{ccccccc}
\scriptsize
\tablecaption{
Summary of IRC~+10216
spectral line data.  The filterbank bandpass was centered at V$_{\rm LSR}$=-40\
kms for all
observations.  Each ratio represents the value of the absorption
feature with respect to nearby stellar continuum at baselines between
2-4 meters.
\label{table:10216data}}
\tablecolumns{7}
\tablehead{
\colhead{Target}        &\colhead{Target}       &
\multicolumn{2}{c}{Bandwidth of} & \colhead{Fringe} &\colhead{ IR}    &
\colhead{Visibility} \\
\colhead{Molecule}&\colhead{Transition}&
\multicolumn{2}{c}{Filterbank} &\colhead{Amplitude} & \colhead{Power} &\colhead
{Ratio} \\
\colhead{}      &\colhead{}     & \colhead{(MHz)}  & \colhead{(km s$^{-1}$)} &
\colhead{ Ratio}
 & \colhead{Ratio} & \colhead{}
}
\startdata
 SiH$_4$& R-(00) A2-A1 & 180 & 1.97 & 0.881$\pm$ 0.013& 0.890$\pm$ 0.006& 0.990
$\pm$ 0.016\\
 SiH$_4$& R-(03) F2-F1 & 180 & 1.86 & 0.912$\pm$ 0.013& 0.907$\pm$ 0.011& 1.006
$\pm$ 0.019\\
 SiH$_4$& R-(03) A1-A2 & 180 & 1.95 & 0.871$\pm$ 0.019& 0.855$\pm$ 0.008& 1.019
$\pm$ 0.024\\
 SiH$_4$& R-(09) A2-A1 & 180 & 1.93 & 0.946$\pm$ 0.031& 0.972$\pm$ 0.007& 0.973
$\pm$ 0.033\\
\hline
 NH$_3$ & aQ(3,3)      & 180 & 1.93 & 0.881$\pm$ 0.024& 0.881$\pm$ 0.020& 1.000
$\pm$ 0.035\\
\enddata
\end{deluxetable}

\subsection{Silane in IRC~+10216}
\subsubsection{Previous work}
\label{section:spectrometer_info}
The mid-infrared transitions of silane around IRC\,+10216 have been
observed by Goldhaber \& Betz (1984),
Goldhaber (1988, hereafter G88), 
Keady \& Ridgway (1993, hereafter KR93), and 
Holler (1999).  Goldhaber and colleagues used a heterodyne
spectrometer which produced the highest spectral resolution data,
fully resolving the absorption line cores (spectral resolution $\sim$
0.2\kms).  KR93 employed a Fourier Transform Spectrometer at Kitt
Peak with somewhat lower spectral resolution, about 3\kms, sufficient
to determine the spectral line strengths but insufficient to fully
resolve the cores.  Holler used a broadband RF spectrometer coupled
to the heterodyne detection system of the ISI, resulting in
sub-\kms spectral resolution (also see Isaak\etal [1999]).   
Because the line depths have remained largely
unchanged over the last 15 years, the modeling results of these
workers was assumed to hold true today.  The most detailed
analysis can be found in KR93, and their gas model parameters were a
starting point for the analysis which follows.

KR93 found a relative abundance of silane compared to molecular
hydrogen of 2.2$\times$10$^{-7}$ with a rotational temperature law,
T$_{\rm rot}= 2000 r^{-0.525}$ (r is expressed in units of stellar radii), 
which falls off slightly faster than
the dust temperature.  The calculated column density was
2.2$\times$10$^{15}$~cm$^{-2}$.  In addition,
T$_{\rm vib}$ was taken
to be about 85\% of the T$_{\rm rot}$ to match the weak 
emission observed in the higher J lines.  These results assumed
spherical symmetry, a uniform outflow of 14\kms outside of 20
R$_\star$, and a microturbulent velocity of 1\kms.  It was found that
the line ratios were much better fit by truncating the silane
distribution inside 40\rstar, and KR93 concluded that silane must be
forming in the outflow at about this radius.  With these assumptions,
KR93 was able to satisfactorily match the line strengths of 7 silane
transitions and the spectral profile of one of Goldhaber's high
spectral resolution profiles.  Holler (1999)
estimated the column density of SiH$_4$ from data taken in May 1999
using one ISI telescope and found it consistent with previous measurements
($\sim$2.7$\times$10$^{15}$ cm$^{-2}$).

\subsubsection{Visibility observations}
Table\,\ref{table:10216data} reports the visibility ratios on and off of 
various silane absorption features in Fall 1998.  A stellar
recessional velocity of V$_{\rm LSR}$=-26.0\kms has been used to
convert V$_{\rm LSR}$ to expansion velocity, based on G88.  In all
cases, the visibility ratio has been observed to be consistent with
unity, and data from all baselines (2-4\,m) have been averaged together.  
Qualitatively, this implies that the absorption region is large
compared to the spatial resolution of the interferometer baseline,
$\sim$0.4\arcsec.

The model for the molecular envelope
developed in KR93 was combined with the new dust shell model from Paper~II
as a starting point for this modeling work.
In particular, the same power law relation
for the rotational temperature was used and the vibrational
temperature was set equal to 85\% of the dust temperature.  A
microturbulent velocity of 0.6\kms was used to match the line widths
observed by Goldhaber (1988).  Next, a series of gas models was
calculated using different silane formation radii: 10\rstar,
40\rstar, and 80\rstar.  The silane abundances were then scaled to
match the average line depths of the observed transitions.  For each of these
3 gas models and for all 4 observed lines, the visibility ratios of the
line compared to the continuum were calculated and compared to
observations.

\begin{figure}[htb]
\begin{center}
\centerline{\epsfxsize=4in{\epsfbox{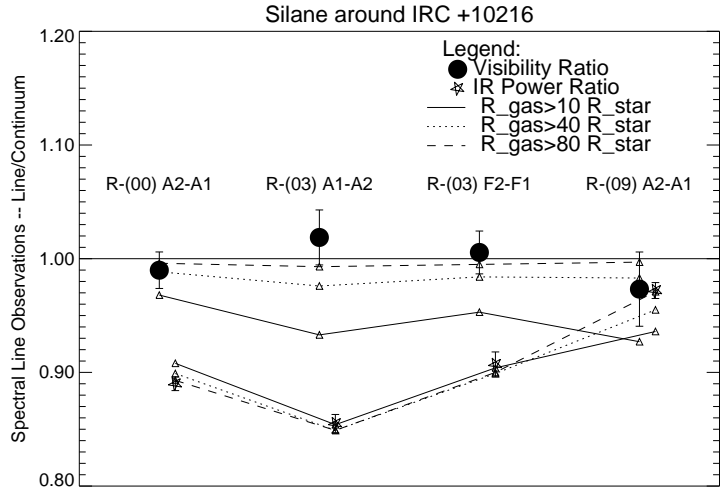}}}
\caption{\scriptsize
This figure
shows the results of modeling the IR power and visibility data for
silane around IRC\,+10216 using the temperature laws derived in
Keady \& Ridgway (1993).  Filterbank data appear as points with error
bars, while model results are represented by small triangles connected
by linestyles corresponding to different molecular density
distributions.  The transitions are presented, from left to right,
in order of increasing excitation energy.
The top points and lines correspond to the visibility
ratios of the absorption line with respect to the continuum, while the
bottom data points and lines correspond to the observed and modelled
line depths.  The solid line is from a model with silane forming at
10\rstar.  The dotted and dashed lines are for molecular formation
radii of 40\rstar\, and 80\rstar\, respectively.
\label{fig:10216_silane_results}}
\end{center}
\end{figure}

The results of these models appear in
figure\,\ref{fig:10216_silane_results}.  Indeed, gas models with
largest formation radii fit the line depths of the four lines the best, as
was found by KR93, implying a relatively low
overall rotational temperature of the absorbing molecules.
In addition, the importance of the visibility ratio
observations made with the ISI and filterbank can be seen.  
For formation radii
less than about 40\rstar, the predicted visibility ratios are
significantly below unity.  Even for a formation radius of 40\rstar\,
(the choice preferred by KR93 based on spectroscopy alone), the 
visibilities in the spectral lines are too small to be consistent with
observations (although
not strongly ruled out).  
The column density for the most favored 80\rstar\, model was found to
1.3$\times$10$^{15}$\,cm$^{-2}$ (2.0$\times$10$^{15}$\,cm$^{-2}$ for
the 40\rstar\, case).

\begin{figure}[hbt]
\begin{center}
\centerline{\epsfxsize=4.5in{\epsfbox{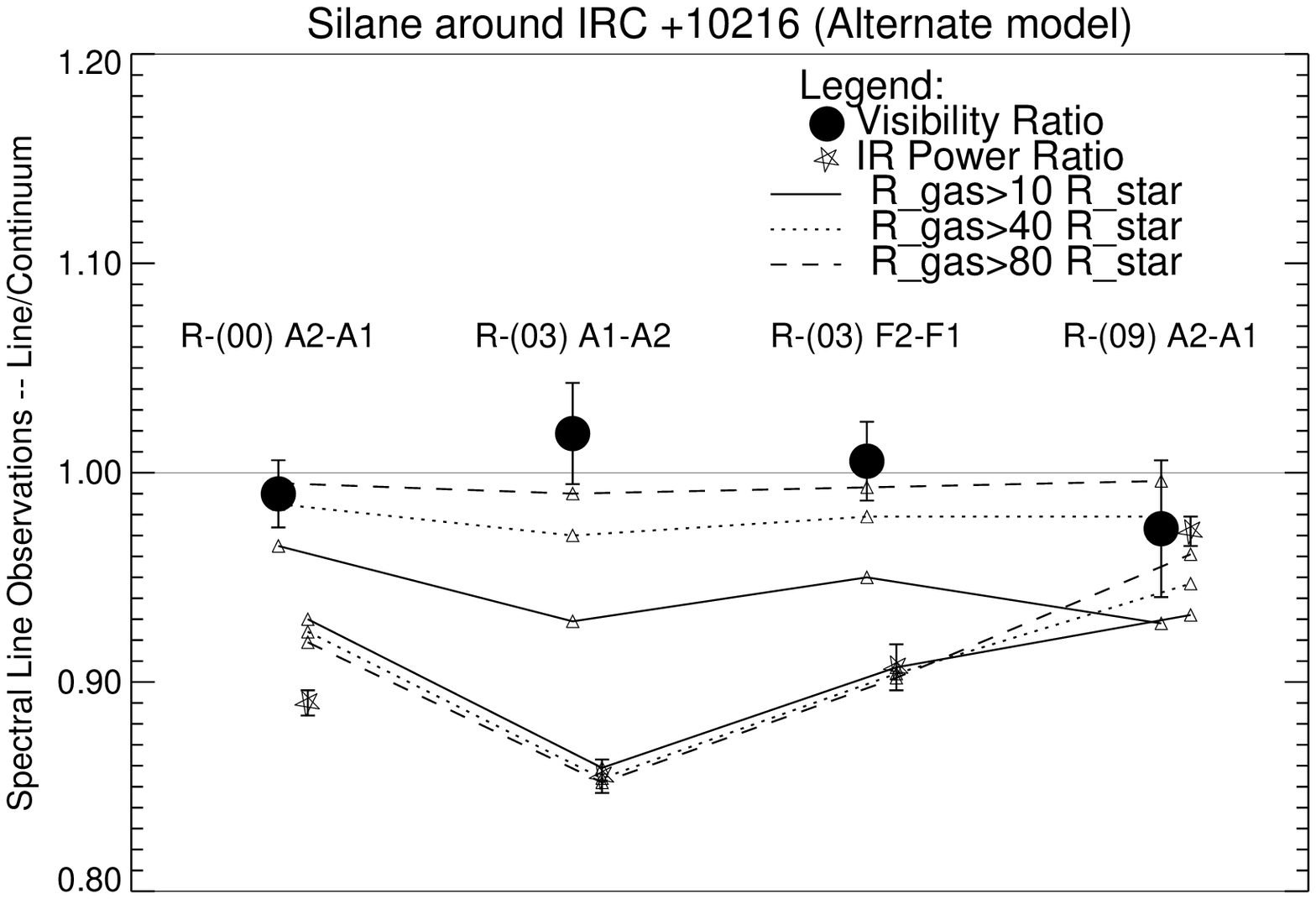}}
}
\caption{
As for
Figure\,{\ref{fig:10216_silane_results}}, but
assuming T$_{\rm rot}$=T$_{\rm dust}$.
\label{fig:10216_silane_alt}}
\end{center}
\end{figure}

Since we have measurements of 4 lines of differing rotational
excitation energies, the sensitivity of the visibility ratios to the
rotational temperature law can be investigated.  Models with identical
gas density distributions were run using T$_{\rm rot}$=T$_{\rm
vib}$=T$_{\rm dust}$, and the results for these calculations can be found in
figure\,\ref{fig:10216_silane_alt}.  This temperature law does not fit
the line depths as well as the first set of models, supporting the
adoption of the empirical temperature fall-off of KR93.
Alternatively, the population of low-lying J transitions could be
enhanced if the mass-loss rate was higher in the past; this would have the
effect of increasing the density of cold silane, increasing the
molecular absorption from the ground state.   The visibility ratios are
similar to those found in the KR93-based models above and the earlier
conclusion of a large molecular formation radius is supported and
now shown to be insensitive to choice of temperature profile.

\subsubsection{Conclusions}
In short, the filterbank observations largely confirm the conclusions
of KR93; namely, that silane must be forming at a large distance from
the star.  
The visibility data directly measures the size of the
absorption region to be larger than $\sim$0.4\arcsec and favors a
formation radius even larger than that proposed by KR93.  The
combination of spectroscopy and interferometry convincingly demonstrates that
silane is forming at a radius of $\simge$80\rstar.  The
visibility data are only able to set a lower limit on the formation radius
since no visibility difference on and off the line was detected
within the experimental uncertainty.

For the stellar radius of 22\,mas used in the model and a distance of
135\,pc, 80\rstar\, corresponds to about 240 AU from the star.  It is
not clear why silane would begin to form at so remote a location in the
outflow.
The densities of the atomic constituents of silane 
are proportional to r$^{-2}$
for a uniformly expanding envelope and would act to shut off
chemical reactions which are density-dependent.  If silane is formed on
the surfaces of grains, then the gas phase density of silicon 
(and its compounds) may not be so important, but rather the
amount of gas-phase silicon and H$_2$ adsorbed onto grains.
Adsorption of H$_2$ on the grain surfaces would probably be small
at the higher temperatures (1000-1400\,K) where dust forms, and be more
pronounced at larger distances form the star (for example, see 
calculations of Leitch-Devlin \& Williams [1985]).
In these models, the
temperatures of the gas and dust are about $\sim$290\,K at 40\rstar
and 200\,K at 80\rstar. 

Gas-phase SiS, which forms in abundant quantities in the photosphere,
has been predicted to be depleted significantly by adsorption onto grains 
(\cite{glassgold92}). In fact, depletion of SiS within 100\rstar\,
has already been observed in both mid-infrared
(\cite{boyle94}) and mm-wave transitions 
(\cite{bieging89}; \cite{bieging93}).  We
hypothesize that the large radius of SiH$_4$ creation is related to
the time-scale for this depletion process and/or the adsorption of H$_2$
onto grains. It is hoped that the introduction of
these new observations will stimulate renewed theoretical interest in
understanding both
the chemical origins and lineage behind the high abundance of silane 
around IRC\,+10216.

\subsection{Ammonia in IRC~+10216}

\subsubsection{Previous work}
The mid-infrared rovibrational
transitions of ammonia around IRC\,+10216 have been
observed by Betz\etal (1979, hereafter B79),
Goldhaber (1988, G88), 
Keady \& Ridgway (1993, KR93), and Holler (1999).  
Some details regarding the spectrometers employed 
have already been summarized in \S\ref{section:spectrometer_info}.
Because the line depths have remained essentially unchanged over the last
15 years, the modeling results of these 
workers are again assumed to be valid today.
As in the case 
for silane, the most detailed analysis
can be found in KR93, and their gas model parameters acted as a starting
point for the analysis which follows.

KR93 found a relative abundance of ammonia compared to molecular
hydrogen of 1.7$\times$10$^{-7}$ with a rotational temperature profile,
T$_{\rm rot}= 2000 r^{-0.60}$ (r is expressed in units of stellar radii),
a faster fall-off than the profile used for silane.
Even with this steeper fall-off,
the predicted depth of the 
lowest lying aR(0,0) line was much weaker than that observed,
requiring either much
colder or more dense gas in the outer envelope ($\simge$400\rstar).
The calculated column density was 2$\times$10$^{15}$ cm$^{-2}$,
very similar to recent results from Holler
($\sim$2.2$\times$10$^{15}$ cm$^{-2}$, \cite{holler99}).  These
results assumed spherical symmetry, a uniform outflow of 14\kms
outside of 20 R$_\star$~, and a microturbulent velocity of 1\kms. 
Since a 
fit of the larger width of the high excitation
aQ(6,6) line with a constant velocity flow was not possible, 
a more complicated
velocity model was used which had the gas accelerating to terminal
velocity at radii 
between 5 and 20\rstar\, (see figures 3a \& 11 in KR93 for the
velocity and density laws adopted).

Although unable to reproduce the line shapes in precise detail, the
KR93 models were successful at qualitatively explaining the bulk of
the spectroscopic observations available by placing the ammonia
density peak around 12\rstar.  Importantly, these models represent 
ideal test cases for the filterbank experiment, because the presence of
ammonia absorption inside of 20\rstar\, is expected to be clearly
indicated by its visibility.

\subsubsection{Results}
Table\,\ref{table:10216data} shows the ratio of the
visibilities observed in and
out of the aQ(3,3) NH$_3$ absorption feature in the Fall of 1998.  A stellar
recessional velocity of V$_{\rm LSR}$=-26.0\kms has been used to
convert V$_{\rm LSR}$ to expansion velocity, based on G88.  The
filterbank bandpass selected ($\sim$1.9\kms) was somewhat smaller than
the full-width half-depth (FWHD) observed by G88, about 5\kms, and
centered on the absorption core, corresponding to an outflow velocity
of 14\kms.  In this case, as for the silane observations, the
visibility ratio was observed to be consistent with unity.

\begin{figure}[htb]
\begin{center}
\centerline{\epsfxsize=3.8in{\epsfbox{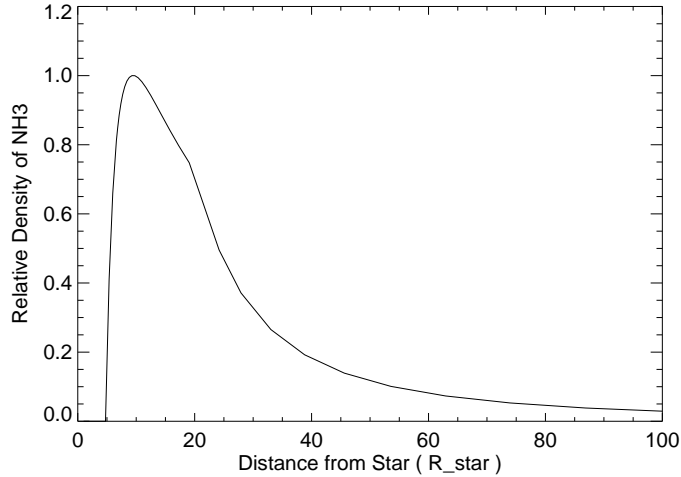}}}
\caption{\scriptsize
This figure
shows the relative density of ammonia as a function of radius for
the IRC~+10216 gas model
(based on Keady \& Ridgway [1993]).
The outflow velocity was increased linearly between 0.1 to 14 \kms
between 3 and 20\rstar.  Using this velocity law and the density of
ammonia shown in the solid line of this figure,
the ammonia shell
mass increases linearly with radius from 5\rstar until 20\rstar, at which
point it becomes constant, consistent with uniform outflow.
The visibility data reported in this paper rules out this model.
\label{fig:nh3_density}}
\end{center}
\end{figure}

The modeling process began with the dust shell model from Paper II and
the gas model developed in KR93.
The same power law relation for the
rotational temperature and a 1.0\kms microturbulent velocity was
initially used.  An ammonia density and velocity distribution similar
to KR93 were used, although simplified.  Specifically, the velocity
was increased linearly from 0.1 to 14\kms between 3 and 20\rstar.  The
ammonia density profile used started at 5\rstar, peaked at
$\sim$10\rstar, and decreased outside of 20\rstar\, consistent with a
uniform outflow ($\rho\propto r^{-2}$).  This corresponded to a linearly
increasing ammonia shell mass until 20\rstar, where uniform outflow was
assumed to begin.  See figure\,\ref{fig:nh3_density} for a plot of the
density profile used.

\begin{figure}[hbt]
\begin{center}
\centerline{\epsfxsize=4.0in{\epsfbox{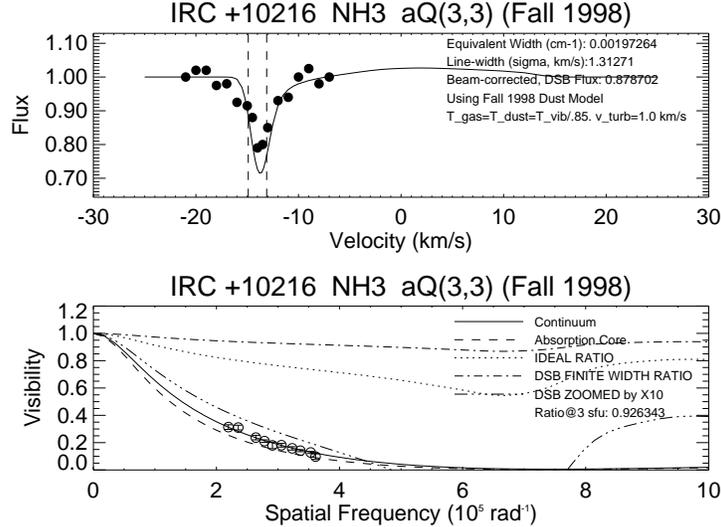}}}
\caption{\scriptsize
This figure shows
the results of modeling the IR power and visibility data for ammonia
around IRC\,+10216 based on parameters derived in Keady \& Ridgway
(1993).  The top panel shows the model spectral line profile
convolved with the instrument response along with the observations of
Goldhaber (1988).  The bottom panel shows the model
visibility curves on and off the spectral feature, and can be compared to
the observed dust continuum data (open circles, see Paper II).
The dashed and solid lines represent the visibility curves on and off the
absorption core (defined by the vertical dashed lines in the top panel)
respectively; the dotted line shows the ratio.
The dot-dashed line shows the visibility ratio after accounting for
finite bandwidth and double-sideband (DSB) detection.  Deviations from
unity are multiplied, or zoomed in, by a factor of 10, and shown with a
3-dots/dashed linestyle; the zoom means that, for example,
0.6 now corresponds to 0.96 in
the original units.
\label{fig:10216_nh3_one}}
\end{center}
\end{figure}

As was shown for silane in the last section, 
significant molecular absorption inside about
20\rstar\,
necessarily results in substantial differences in
interferometric visibilities
on and off 
the spectral line.  Figure\,\ref{fig:10216_nh3_one} shows the spectral
profile and visibility ratio for the above model of ammonia around
IRC~+10216, based on KR93, and also the spectral line data taken from
G88.  KR93 did not model this specific transition, so 
the agreement can not be expected to be precise; indeed, the line prediction
is narrower
than the profiles observed by Goldhaber.  The density of ammonia has
been adjusted to match the line depth observed in Fall 1998 within the
bandpass of the ISI filterbank, resulting in a column density of
8.4$\times$10$^{15}$\,cm$^{-2}$ for this model.  

The model visibility ratio at a spatial frequency of 
$3 \times 10^5$ radians$^{-1}$ is 
0.926 compared to the observed ratio of 1.000$\pm$0.035.  This
model is thus ruled out at a 2-$\sigma$ level.  This reinforces a
conclusion from the modeling of silane: significant molecular
absorption occurring within about 20\rstar\, for IRC\,+10216 results in a
strong on-line and off-line difference in the interferometric
visibilities, a signal not seen with the
ISI filterbank experiment.  The conclusions that ammonia forms at a distance
of $\simge$20\rstar from the star indicates that, as in the case of silane,
ammonia molecules do not form near where the dust condenses but rather at 
lower temperatures and probably through surface adsorption.  The dust grain
temperature at 20\rstar is $\sim$400\,K in these models.

\subsubsection{Discussion}
\label{section:10216_nh3_discussion}
There are other, purely spectroscopic, observations which also suggest
that the explanation put forth by KR93 for the broad aQ(6,6) line,
bulk acceleration of gas within 20\rstar, may not be
fully correct.  KR93 did not attempt to
fit the high spectral resolution profiles
published in G88, which revealed details concerning the
ammonia around IRC\,+10216 not reproduced by models relying on
accelerating gas to broaden the absorption lines.

KR93 fit the aQ(6,6) line data originally published in B79 which was
cut-off at -17\kms relative to the stellar velocity.  Features
broadened due to absorption by gas in the acceleration region would
be asymmetric, with a sharp blue edge and significant
red-shifted absorption by hot gas.  Indeed, the model profiles of KR93
show this telltale characteristic (see figure\,13 of KR93), but were
unconstrained by the incomplete B79 published profile.  Complete line profiles
including the full blue-wing of the absorption feature were published
in G88, revealing a rather symmetric line -- at odds with KR93 model
results.

Thus far, the linewidth data from G88 can not be reproduced in detail with
this class of models, independent of the visibility data.  The line
widths were observed to increase with the energy of the rotational
level.  The 1-$\sigma$ widths were measured to be 0.8\kms, 1.7\kms,
and 3.1\kms for the aR(0,0), aQ(2,2), and aQ(6,6) transitions
respectively, while the relative velocity of the core remained
constant to within $\sim$1\kms.  While a near-star
acceleration law and suitably peaked ammonia distribution can
be adjusted to
reasonably fit the linewidth of any single transition
(e.g., figure\,\ref{fig:10216_nh3_one}), the large increase between the
aQ(2,2) and aQ(6,6) line is not reproduced.  Furthermore, lines broadened
by accelerating gas should show significant shifts in the location of the core
as a function of excitation energy.

The presence of decaying gas turbulence outside the acceleration region
may explain qualitatively
the linewidth behavior and the lack of a
significant difference in the visibilities. 
The inner dust shell of IRC +10216 is already known to be
quite clumpy in near-infrared images 
(\cite{weigelt98a,haniff98,tuthill98b,mythesis,tuthill2000}).
Furthermore, dynamical models of dust shell
production in carbon stars lead to large velocity dispersion as the
individual shells of dust are accelerated
(\cite{winters95,weigelt98a,winters98}).
Both turbulence and velocity gradients
across individual dust shells (clumps) will cause significant
broadening of lines formed in these regions.  Under these conditions, one 
would also
expect the doppler velocity of the core to be nearly constant,
consistent with observations, but the
line widths would naturally decrease for lines forming further in the
outflow as sound waves (or weak shocks) bring the outflow into
hydrostatic equilibrium.  The dynamical
dust shell models of Winters\etal could be coupled 
with the radiative transfer code of
Keady to quantitatively test these suppositions
(as for CO in Winters\etal [1998]).

There are other potential explanations for why  
absorption lines by high-J states
are much broader than for lower-excitation states.
For instance, if the high-J states are populated by shock-heating, 
one would expect a higher velocity dispersion due the turbulence and
velocity shifts accompanying shocks.  This class of explanation, 
whereby the excitation mechanism (and possibly even the {\em formation} 
mechanism) is directly related to peculiar {\em local} conditions
(e.g., shocks, high velocity dispersion), needs further consideration.  

\begin{figure}[htb]
\begin{center}
\centerline{\epsfxsize=\columnwidth{\epsfbox{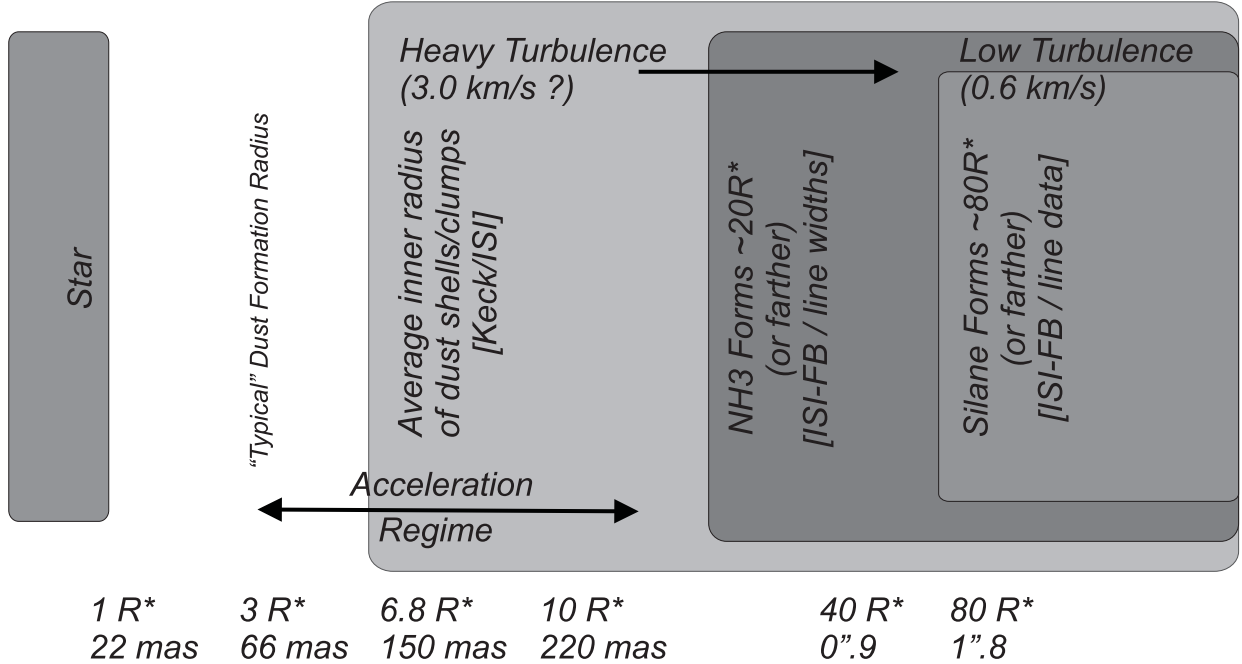}}}
\caption{\scriptsize
This figure
illustrates the likely sites of silane and ammonia formation around IRC~+10216
based on the results of this paper and previous spectral line work.
Importantly, the data suggests that
the outflow remains quite turbulent and/or clumpy significantly beyond the
acceleration region.  The observations undergirding each major
element of this schematic are listed in brackets
(Keck = near-IR speckle work [Tuthill\etal 2000],
ISI = ISI continuum data [Paper II], ISI-FB = ISI filterbank
data [this work]).
\label{fig:10216_molecules}}
\end{center}
\end{figure}

\subsection{Summary for IRC~+10216}
The narrow line widths of all the silane lines and the lack of an
interferometric signal place the location of silane formation at or beyond 
$\sim$80\rstar\, outside the turbulent inner envelope.  
In addition, the
lack of interferometric signal and the broad (symmetric) lines of
ammonia place their formation outside of $\sim$20\rstar, beyond the gas
acceleration region, but in a turbulent (or clumpy) flow.  These
ideas are illustrated schematically in figure\,\ref{fig:10216_molecules}
and represent a significant improvement to our knowledge of the
molecular stratification around IRC~+10216.  

\section{VY CMa}
\label{section:vycma_line_results}
While no silane has been observed in the oxygen-rich stellar wind of 
the red supergiant VY~CMa, many absorption lines of ammonia have.
A journal of these spectral line observations made with the ISI
filterbank can be found in table\,\ref{table:vycmajournal}.

\begin{deluxetable}{cccccl}
\scriptsize
\tablecaption{
Journal of VY~CMa spectral line observations
\label{table:vycmajournal}}
\tablehead{
\colhead{Target}        &\colhead{Target}       &
\colhead{LO}    &\colhead{LO}   &\colhead{LO} & \colhead{Dates}\\
\colhead{Molecule}&\colhead{Transition}&\colhead{Molecule}&
\colhead{Transition}&\colhead{Wavelength} & \colhead{(U.T.)}\\
         &          &        &          &\colhead{($\mu$m)} & }
\startdata
NH$_3$ & aQ(2,2) & $^{13}$C$^{16}$O$_2$ & R(24) & 10.738 &
1998 Oct 20, Oct 21, Oct 22, Nov 25 \\
NH$_3$ & sP(4,3) & $^{13}$C$^{16}$O$_2$ & P(30) & 11.262 &
1998 Oct 27 \\
NH$_3$ & aQ(6,6) & $^{13}$C$^{16}$O$_2$ & R(18) & 10.784 &
1998 Oct 18, Oct 19, Oct 28, Nov 05, Nov 24, Dec 11\\
\enddata
\end{deluxetable}

\begin{deluxetable}{ccccccc}
\tablecaption
{Summary of VY~CMa
spectral line data.  The filterbank bandpass was 
centered at V$_{\rm LSR}$=-7\kms for all
observations, except when observing the red-half 
of aQ(6,6) (V$_{\rm LSR}$=-6\kms).
Each ratio represents the value of the absorption
feature with respect to nearby stellar continuum at baselines between
2-4 meters.
\label{table:vycmadata}}
\scriptsize
\tablecolumns{7}
\tablehead{
Target  &Target & \multicolumn{2}{c}{Bandwidth of} & Fringe & IR    &  Visibility \\
Molecule&Transition& \multicolumn{2}{c}{Filterbank} &Amplitude & Power & Ratio
\\
        &       & (MHz)  & (km s$^{-1}$) & Ratio & Ratio &
}
\startdata
 NH$_3$ & aQ(2,2)      & 540 & 5.80 & 0.933$\pm$0.022& 0.932$\pm$ 0.012& 1.001$
\pm$ 0.027\\
 NH$_3$ & sP(4,3)      & 420 & 4.73 & N/A  & 0.983$\pm$ 0.026&  N/A \\
 NH$_3$ & aQ(6,6)      & 540 & 5.82 & 0.922$\pm$ 0.024& 0.950$\pm$ 0.012& 0.971
$\pm$ 0.028\\
 NH$_3$ & aQ(6,6)      & 180 & 1.94 & 0.958$\pm$ 0.080& 0.956$\pm$ 0.046& 1.002
$\pm$ 0.097\\
        & Red Half     &     &      & && \\
\enddata
\end{deluxetable}

\subsubsection{Previous work}
The mid-infrared transitions of ammonia around VY~CMa were first
observed by McLaren \& Betz (1980).  However,
a larger set of data with better frequency coverage was published by
Goldhaber (1988, G88) and these data have been used for
subsequent consideration.  Details regarding the capabilities of the
heterodyne spectrometer employed can be found in
\S\ref{section:spectrometer_info}.  While McLaren \& Betz 
reported a significant
change in the line profile of NH$_3$ aQ(2,2) between October 1978 and
December 1979, the profile reported by G88 observed in September 1982
is nearly identical to the December 1979 data.  In addition, the G88
line depth is consistent with the spectral data obtained by the
filterbank in this paper (see table\,\ref{table:vycmadata}).  In
light of this consistency, the assumption that the G88 line
profiles are still representative of the current epoch has been made.

All of the lines, aR(0,0), aQ(2,2), aQ(3,3), \& aQ(6,6), observed by
G88 possessed broad absorption features with 1-$\sigma$ widths of
$\sim$3.8\kms.  The cores of the lines were centered around V$_{\rm
LSR}\sim -6$\kms, with the high excitation line cores being slightly
less blue-shifted.  Assuming a stellar recessional velocity of V$_{\rm
LSR}=22.3$\kms (based on maser data), this corresponds to an
outflow rate of $\sim$29\kms.  Using these lines, G88 estimated a
column density of ammonia of (6$\pm$4)$\times$10$^{15}$ cm$^{-2}$
using a simple model based on the Sobolev approximation.  We
have repeated these calculations using a similar gas model with the Keady
code.

\subsubsection{Results and models}
\label{section:jkstates_nh3}
Table\,\ref{table:vycmadata} shows the visibility ratios observed on
and off of a number of NH$_3$ absorption features in Fall 1998.  A
stellar recessional velocity of V$_{\rm LSR}$=22.3\kms has been used
to convert V$_{\rm LSR}$ to expansion velocity, based on G88.  The
filterbank bandpasses selected were set to equal the full-width
half-depths (FWHD) observed by G88, and were centered on the absorption
cores.  Data in table~\ref{table:vycmadata} represents a
non-detection of the sP(4,3) line, which also had not been detected by any
previous measurements.
Just as
for the IRC\,+10216 (see table\,\ref{table:10216data}), the visibility
ratios were observed to be consistent with unity.  In addition to the
full aQ(6,6) absorption line, the red half of the line was observed
separately in the hopes of observing the location of any accelerating gas.
However, the SNR for this measurement ($\sim$10) was not high enough
to justify further modeling.

A simple gas model based on the line-fit of the aR(0,0) line in G88
was developed using the new VY~CMa dust shell model of
Paper~II.  Unlike the case for IRC\,+10216, the
linewidths for the NH$_3$ lines around VY\,CMa were all roughly the
same.  This fact and that the line core locations only showed a slight
shift with excitation energy support a model in which the absorption
lines all form mostly outside of the acceleration region.  Therefore a
constant outflow model was adopted using an expansion velocity of
29\kms, which matched the profiles of the aQ(2,2) and aQ(6,6) lines;
the broad linewidths were created by model microturbulence of 3.5\kms.

The temperature of the J=K rotational states is controlled either by
collisions, due to the lack of dipole transitions from these states to
lower J, or by IR-pumping in the 6.1\,$\mu$m band (see
\S\ref{section:gastemps}).  During modeling, 
this temperature profile
is normally varied using a power-law (as by KR93) to empirically fit
the line depth ratios.  Unfortunately there were insufficient number of
lines available in this case to utilize this strategy.  An 
alternative scheme, 
in which two temperature profiles were assumed to 
represent reasonable limits, was used here:

\begin{enumerate}
\item{The rotational temperature was set equal to the dust temperature, 
   T$_{\rm rot}$=T$_{\rm dust}$.  This corresponded to full coupling of the
internal degrees of freedom of the gas molecules to the radiation field.}
\item{
  T$_{\rm rot}=$T$_\star  r^{-0.60}$, where r is expressed in units of stellar radii.
This choice is
similar to that found empirically for ammonia around IRC +10216 by 
Keady \& Ridgway (1993).  While VY~CMa and IRC\,+10216 do have similar 
mass-loss
rates (within a factor of $\sim$10), the differing chemistry (O-rich vs. C-rich) may affect the
cooling rates, and this ad hoc temperature profile is clearly 
rather speculative. }
\end{enumerate}

Because G88 in most cases did not publish the emission 
components of the NH$_3$
transitions, we were not able to model this
aspect of the line profile.  Instead we followed KR93 (for
IRC\,+10216) and suppressed emission in our model profiles by
assuming the vibrational temperature was 85\% of the dust temperature.

\begin{figure}[htb]
\begin{center}
\centerline{\epsfxsize=4.0in{\epsfbox{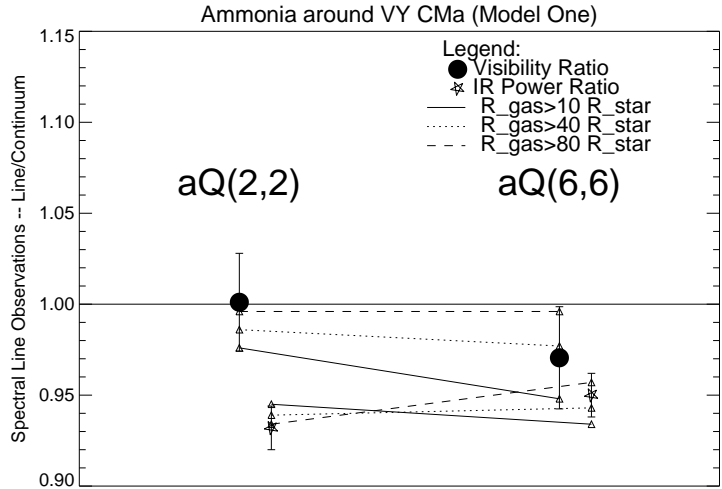}}}
\caption{\scriptsize
This figure
shows the results of modeling the IR power and visibility data for ammonia 
around
VY~CMa assuming T$_{\rm rot}$=T$_{\rm dust}$.  Filterbank data appear as 
points with
error bars, while model results  are represented by small triangles 
connected by
linestyles corresponding to different molecular density distributions.
The top points and lines correspond to the visibility ratios of the
absorption line with respect to the continuum, while the bottom data points 
and lines
correspond to the observed and modelled line depths.  The solid line is 
from a model
with silane forming at 10\rstar.  The dotted and dashed lines are for molecular
formation radii of 40\rstar\, and 80\rstar\, respectively.
\label{fig:vycma_nh3_one}}
\end{center}
\end{figure}

\begin{figure}[hbt]
\begin{center}
\centerline{\epsfxsize=4in{\epsfbox{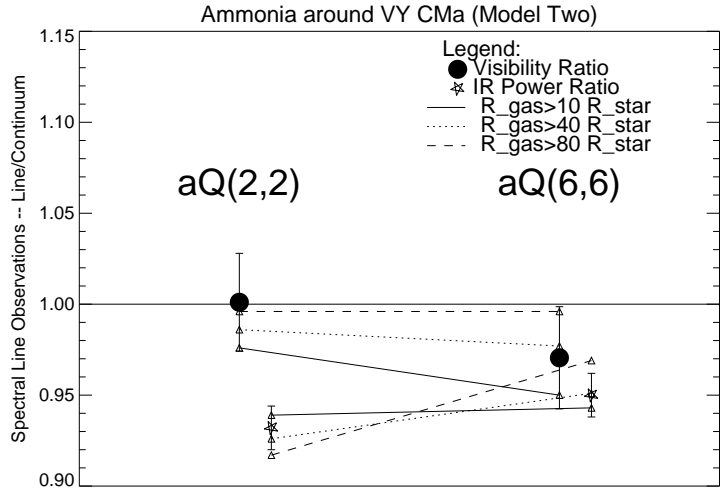}}}
\caption
{\scriptsize Similar to
Figure\,\ref{fig:vycma_nh3_one}, but assuming T$_{\rm rot}=$T$_\star
r^{-0.60}$, where r is in units of stellar radii.
\label{fig:vycma_nh3_two}}
\end{center}
\end{figure}

Figure\,\ref{fig:vycma_nh3_one} shows the results assuming T$_{\rm
rot}$=T$_{\rm dust}$.  The high optical depth at 11\,$\mu$m
($\tau\sim2.4$, see Paper~II) effectively shields the inner dust and
molecular envelope from view.  The $\tau_{\rm dust}$=1 surface is
located at about 12\rstar\, along the central impact parameter and
hence most of the observed continuum emission is occurring outside of
this.  This explains why none of the models with molecular formation
radii between 10-80\rstar\, were decisively ruled out with the
interferometry data of figure\,\ref{fig:vycma_nh3_one}.  However, one
can see that for this assumed temperature law, the line ratios were
best fit by a molecular formation radius outside of $\sim$40\rstar.
Interestingly, this distance corresponds to the extent of the
acceleration region deduced from proper motion studies of the H$_2$O
masers (\cite{marvel96}; \cite{richards98}).
Figure\,\ref{fig:vycma_nh3_two} shows the results for the temperature
profile of case 2, assuming T$_{\rm rot}=$T$_\star r^{-0.60}$; the
same general conclusions still apply, showing that they are not highly
model-dependent.  It is worth noting that the equivalent widths for
models with small molecular formation radii (especially for 10\rstar)
were systematically higher than those measured by G88.  While the line
depths in the filterbank bandpass were fitted by increasing the
ammonia density, the linewidths themselves were not.  Ideally, the
microturbulence parameter should be adjusted for each set of molecular
formation radii to compensate for differing line broadening effects,
such as from thermal broadening and the finite size of the continuum
source.  Such adjustments did indeed improve the quality of these
fits, but significantly increased the complexity of the modeling while
only weakly affecting the line and visibility ratios.

\subsubsection{Discussion}
We conclude that NH$_3$ probably 
forms near the termination of the acceleration phase
in the circumstellar envelope of VY~CMa ($\sim$40\rstar).  This hypothesis
is supported by the following observations:
\begin{itemize}
\item{Weak (but non-negligible) correlation of absorption core velocity with
excitation energy suggests ammonia exists to a limited degree in the
acceleration region.}
\item{Line ratios of the aQ(2,2) and aQ(6,6) transitions were best fitted
by a molecular formation radii $\simge$40\rstar, a result 
found to be insensitive to
the rotational temperature profile.}
\item{The visibility data for the high excitation 
aQ(6,6) line marginally supports a large 
molecular formation radius, but the high optical depth of the
dust shell makes the interferometric data relatively insensitive to absorption
in the inner circumstellar envelope.}
\end{itemize}

Around IRC\,+10216, high excitation NH$_3$ lines were observed to be
significantly broader than for the low-lying J-states.  Interestingly,
this behavior was {\em not} duplicated for NH$_3$ around VY~CMa.
Unlike the case
for IRC\,+10216, it appears that ammonia is forming at least
partly in the acceleration region around this star and hence the line
formation of all the transitions is more affected by large turbulence
in this region.  Alternatively, the highly asymmetric inner dust shell
seen in the near-infrared (\cite{monnier99})
and hinted at by ISI visibility data (Paper~II) 
may be influencing the line formation
characteristics.  If the outflow is significantly asymmetric,
linewidths will be clearly affected by the varying projected outflow
velocities of absorbing gas.  Continued monitoring of water masers in
the inner dust shell may clarify the gas dynamics and level of gas
turbulence.  This information will be critical for further progress in
understanding the geometry of the molecular envelope around VY~CMa.

\section{Conclusions}
We have presented the first interferometric results on mid-infrared
spectral lines around evolved stars.  Specific results for the
carbon star IRC~+10216 and the red superigant
VY~CMa have been summarized in previous sections.  In addition, a
few general conclusions can be drawn:

\begin{itemize}
\item{The formation radii of silane and ammonia are significantly beyond
the dust formation zone for both evolved stars examined here.}
\item{Since dust formation by itself does not catalyze formation of these
molecules, some other physical mechanism(s), still unknown,
must be at play.  Probably the adsorption of gas-phase molecules
(e.g., SiS or H$_2$) onto grains set the time/length-scale for 
the important chemical reactions occurring on the grain surfaces.}
\item{When coupled with spectroscopic observations, these results 
indicate that turbulence, or velocity dispersion, 
is quite high in the inner envelope and is more likely
responsible for the broad linewidths of high excitation transitions 
than bulk acceleration.}
\end{itemize}

\acknowledgements
{
The authors would like to recognize W. Fitelson, C. Lionberger, and
M. Bester for important contributions to the construction and software
development of the filterbank, and K. McElroy for observing 
assistance.
We also acknowledge productive discussions with
A. Glassgold and A. Betz.   JDM also would like to thank 
J. Keady for kindly allowing his sophisticated radiative
transfer code to be used for this project, and
for many stimulating discussions about the molecules of IRC~+10216.
This work is part of a
long-standing interferometry program at U.C. Berkeley, supported by
the National Science Foundation (Grant AST-9221105, AST-9321289, and 
AST-9731625)
and by the Office of Naval Research (OCNR N00014-89-J-1583).}

\end{document}